\documentclass[10pt,conference]{IEEEtran}
\usepackage{epsfig,rotating,setspace,latexsym,amsmath,epsf,amssymb,amsfonts,bm,theorem,cite,authblk, bbm,color, multirow, caption, subcaption}
\IEEEoverridecommandlockouts
\allowdisplaybreaks

\title{Joint Sensing and Task-Oriented Communications with Image and Wireless Data Modalities for Dynamic Spectrum Access}
\begin{document}
\author[1]{Yalin E. Sagduyu}
\author[1]{Tugba Erpek}
\author[2]{Aylin Yener}
\author[3]{Sennur Ulukus}

\affil[1]{\normalsize  Virginia Tech, Arlington, VA, USA}

\affil[2]{\normalsize  The Ohio State University, Columbus, OH, USA}

\affil[3]{\normalsize University of Maryland, College Park, MD, USA}
\maketitle

\begin{abstract}
This paper introduces a deep learning approach to dynamic spectrum access, leveraging the synergy of multi-modal image and spectrum data for the identification of potential transmitters. We consider an edge device equipped with a camera that is taking images of potential objects such as vehicles that may harbor transmitters. Recognizing the computational constraints and trust issues associated with on-device computation, we propose a collaborative system wherein the edge device communicates selectively processed information to a trusted receiver acting as a fusion center, where a decision is made to identify whether a potential transmitter is present, or not. To achieve this, we employ task-oriented communications, utilizing an encoder at the transmitter for joint source coding, channel coding, and modulation. This architecture efficiently transmits essential information of reduced dimension for object classification. Simultaneously, the transmitted signals may reflect off objects and return to the transmitter, allowing for the collection of target sensing data. Then the collected sensing data undergoes a second round of encoding at the transmitter, with the reduced-dimensional information communicated back to the fusion center through task-oriented communications. On the receiver side, a decoder performs the task of identifying a transmitter by fusing data received through joint sensing and task-oriented communications. The two encoders at the transmitter and the decoder at the receiver are jointly trained, enabling a seamless integration of image classification and wireless signal detection. Using AWGN and Rayleigh channel models, we demonstrate the effectiveness of the proposed approach, showcasing high accuracy in transmitter identification across diverse channel conditions while sustaining low latency in decision making. 
\end{abstract}

\section{Introduction} \label{sec:Intro} 
The increasing demand for wireless communication services has led to a spectrum scarcity challenge, highlighting the need for efficient spectrum management strategies. Dynamic Spectrum Access (DSA) has emerged as a promising solution to optimize spectrum utilization by allowing unlicensed users to opportunistically access underutilized frequency bands. However, efficient implementation of DSA requires robust spectrum sensing techniques to identify and avoid interference with incumbent users. In this context, the integration of diverse modalities, such as image and wireless (spectrum) data, presents a novel and powerful approach to enhance target  spectrum sensing capabilities.

\begin{figure}[h!]
	\centering
	\includegraphics[width=\columnwidth]{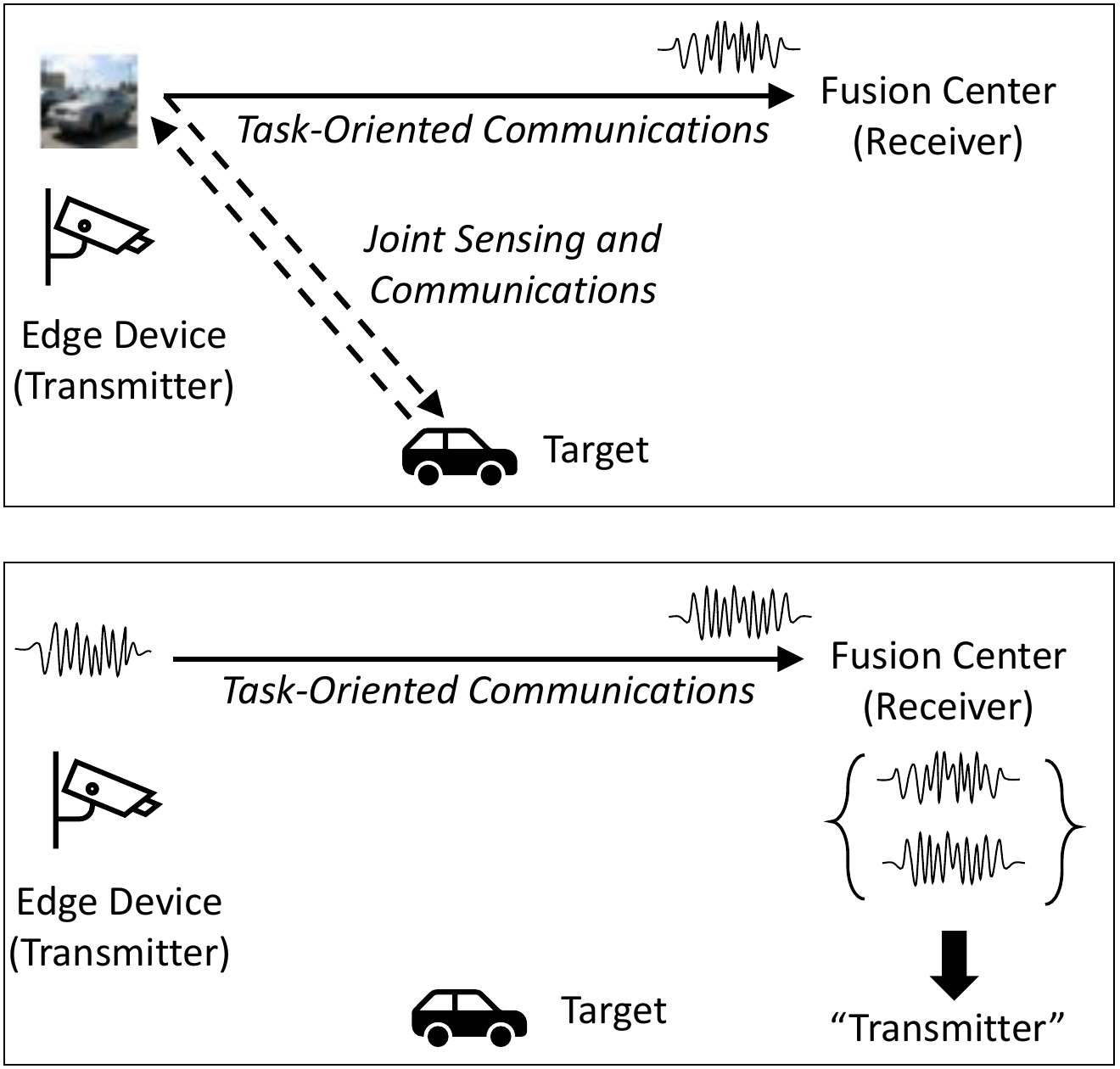}
	\caption{Multi-modal data collection and delivery to assist with DSA.}
	\label{fig:scenario}
 \vspace{0.2cm}
\end{figure}

Traditional spectrum sensing methods often rely solely on signal processing and/or machine learning techniques using spectrum sensing data, which may be limited in their ability to discriminate between different types of transmitters and the environmental context in which they operate. Recognizing the potential of \emph{multi-modal data} for comprehensive situational awareness, this paper introduces a deep learning approach that synergistically leverages both image and target
sensing data for the DSA purposes, as illustrated in Fig.~\ref{fig:scenario}.

The proposed approach centers around an edge device equipped with a camera that collects image samples for object recognition, particularly focusing on identifying potential transmitters, such as those positioned at vehicles. Acknowledging the communication and computational constraints and trust concerns associated with on-device computation, our system introduces a collaborative framework. This involves the selective transmission of processed information of much lower dimension from the edge device to a trusted receiver, functioning as a fusion center where the decision of transmitter identification is made. 

To streamline the transmission process and address the challenge of conveying critical information within limited bandwidth, we employ the concept of \emph{task-oriented communications} that optimizes the use of limited resources by tailoring the transmission to the specific goal of the receiver, avoiding unnecessary data transmission and reconstruction. This resource-efficient approach is particularly advantageous in DSA scenarios, where quick and accurate decisions are imperative, and communication and computational resources are often constrained. For that purpose, we are employing an \emph{encoder} at the transmitter to perform joint source coding, channel coding, and modulation, efficiently transmitting essential information of reduced dimension specifically tailored for object classification. This way, the receiver can reliably complete its task (i.e., make image classification decision) without reconstructing data samples. 

In the meantime, the transmitted signals may interact with the environment and return to the transmitter, allowing for the collection of target sensing data. Traditionally, target sensing systems often resort to the transmission of dedicated probing signals to gather information about the spectral environment. This practice, while effective, comes at the cost of additional bandwidth consumption and potential interference with existing communication channels. Moreover, the deployment of extra probing signals may introduce delays and increase the complexity of the overall system.

Recognizing these challenges, our proposed approach is based on \emph{joint (integrated) sensing and communications}. By combining target sensing and communication tasks within a unified framework, our system leverages the returned signals from the environment, originally transmitted for object recognition, as a valuable source of sensing data. This eliminates the necessity for additional probing signals, thereby conserving precious bandwidth resources and reducing the latency associated with target sensing.

The collected target sensing data undergoes a second round of \emph{encoding} at the transmitter, potentially condensing the information into a reduced-dimensional format for communication back to the fusion center through task-oriented communications. On the receiver side, a \emph{decoder} plays a pivotal role in identifying active transmitters by fusing data received through both image transmission and target sensing. Importantly, the proposed system jointly trains the two encoders at the transmitter and the decoder at the receiver corresponding to three deep neural networks (DNNs). This approach enables a seamless integration of multi-modal data for image classification and wireless signal detection to identify potential transmitters for DSA. Note that this approach identifies the presence of transmitters even when they are passive (i.e., they are not transmitting) or their transmissions do not reach the spectrum sensors (e.g., because they are outside the transmission range or employ directional transmissions so that they cannot be captured by spectrum sensors) although they may reach and interfere with the receiver to be protected (such as the primary user receiver). In this context, the proposed approach aims to assist conventional spectrum sensing techniques that are employed to identity active transmissions.   

To validate the efficacy of the proposed approach, we leverage the CIFAR-10 dataset as our source of images, creating a realistic representation of objects, including potential transmitters. Furthermore, we simulate communication and sensing in diverse channel conditions by incorporating Additive White Gaussian Noise (AWGN) and Rayleigh channel models. The results showcase high accuracy in transmitter identification across a spectrum of challenging channel conditions, while maintaining low latency for practical applicability in real-world DSA scenarios such as in the Citizens Broadband Radio Service (CBRS) band, where prompt and accurate decision-making is crucial for efficient spectrum utilization.

The remainder of the paper is organized as follows. Sec.~\ref{sec:related_work} discusses related work. Sec.~\ref{sec:system_model} presents the joint sensing and task-oriented communications approach. Sec.~\ref{sec:DL} describes the deep learning framework. Sec.~\ref{sec:perf} presents the performance evaluation results. Sec.~\ref{sec:Conclusion} concludes the paper. 

\section{Related Work} \label{sec:related_work} 
By monitoring and analyzing the spectrum, DSA systems can identify available frequency bands and adapt in real-time to optimize spectrum utilization. This data-driven approach enhances the efficiency of wireless communications, allowing for agile and adaptive allocation of frequencies based on current environmental conditions and usage patterns.

Integrating multiple data modalities, including spectrum data and images captured by edge devices, is expected to enhance the capabilities of DSA. Recently, image data has found ways to assist wireless tasks such as channel estimation and beam tracking \cite{charan2022computer, alkhateeb2023deepsense}. When applied to DSA, the multi-modal data processing approach can be effectively used to  authenticate devices or facilitate the recognition of primary users, allowing the DSA system to prioritize and allocate spectrum resources based on the characteristics and requirements of authorized users, thereby optimizing overall spectrum utilization.

One challenge is how to deliver multi-modal image and target sensing data from the edge devices to the receiver that needs to perform the task of transmitter identification. Note that edge devices have limited communication and computational resources, and cannot be trusted to perform the task of transmitter identification. The primary goal of conventional communications has been the reliable delivery of information over a channel by minimizing some form of reconstruction loss such as the mean squared error. Recently, semantic communications has emerged to preserve semantics (meaning) of information during the information delivery \cite{guler2014semantic, guler2018semantic, gunduz2022beyond, chaccour2022less, xie2021deep, qin2021semantic, uysal2021semantic}. Specifically, the preservation of meaning can be achieved by ensuring that the received information leads to the correct outcome of a machine learning task \cite{sagduyu2022semantic}. 

To that end, semantics can be represented by the importance of the task to be completed, leading to the concept of task-oriented (or goal-oriented) communications, where the objective is to complete a task (e.g., a machine learning task) at a receiver by using the data samples that are originally available at the transmitter \cite{zhang2022deep, sagduyu2023task, sagduyu2023age, shao2021learning, strinati20216g, kang2022task, qin2022timeliness, shi2023toc}. In contrast to the conventional approach of transmitting and reconstructing all data samples, task-oriented communications delivers only a reduced amount of information to the receiver for the purpose of completing the underlying task as reliably as possible. This approach makes efficient use of channel resources and incurs low latency by reducing the dimension of information delivered over the channel. 

Both semantic and task-oriented communications find common ground in the application of deep learning. To this end, a communication system akin to autoencoder communications \cite{Oshea1, erpek2020deep}  is established, involving the joint training of an encoder at the transmitter and a decoder at the receiver—both implemented as DNNs. In traditional communications systems, the transmitter and receiver functionalities are separated in design and modularized into individual communication blocks that are typically designed by analytical models. An autoencoder that consists of an encoder at the transmitter and a decoder at the receiver  can be trained by taking the channel and hardware impairments into account. In this context, the encoder represents channel coding and modulation operations and the decoder represents demodulation and channel decoding operations. Starting with source data samples instead of bits or symbols, this setup is extended to the encoder conducting joint source coding, channel coding, and modulation operations on input samples, such as images. For semantic communications, the decoder engages in joint demodulation, channel decoding, and source decoding operations on received signals to reconstruct data samples while simultaneously performing a task on the received signals. On the other hand, in task-oriented communications, the focus is solely on executing tasks on the received signals without the necessity to reconstruct data samples. Task-oriented communications can be designed to serve multiple users each with different data modalities  \cite{xie2021task, xie2022task}

Delivery of data such as images from the edge devices to the fusion center can also serve the dual purpose of target sensing. 
Joint (or integrated) sensing and communications refer to the integration of sensing (data collection or perception) and communication functionalities within a unified system \cite{demirhan2022integrated, liu2022integrated, xiong2023, cui2021integrating, du2022integrated, liu2022learning, Zhang2021AnOO}. Inclusion of radar-like sensing capabilities in communication networks is desired in next-generation wireless systems to utilize the spectrum more efficiently while improving the performance. This approach enables a system to not only gather information from the environment through sensors but also to communicate this information effectively to other components or system such as in the Internet of Things (IoT) and sensor networks The synergy between sensing and communication leads to real-time responsiveness, energy efficiency, and enhanced information fusion. Joint sensing and communications can be setup in forms of autoencoder communications \cite{mateosae2022, muth2023} and can be integrated with semantic and task-oriented communications, as discussed in \cite{sagduyu2023joint}.

\section{Joint Sensing and Task-Oriented Communications} \label{sec:system_model}
\begin{figure*}[h]
	\centering
	\includegraphics[width=1.7\columnwidth]{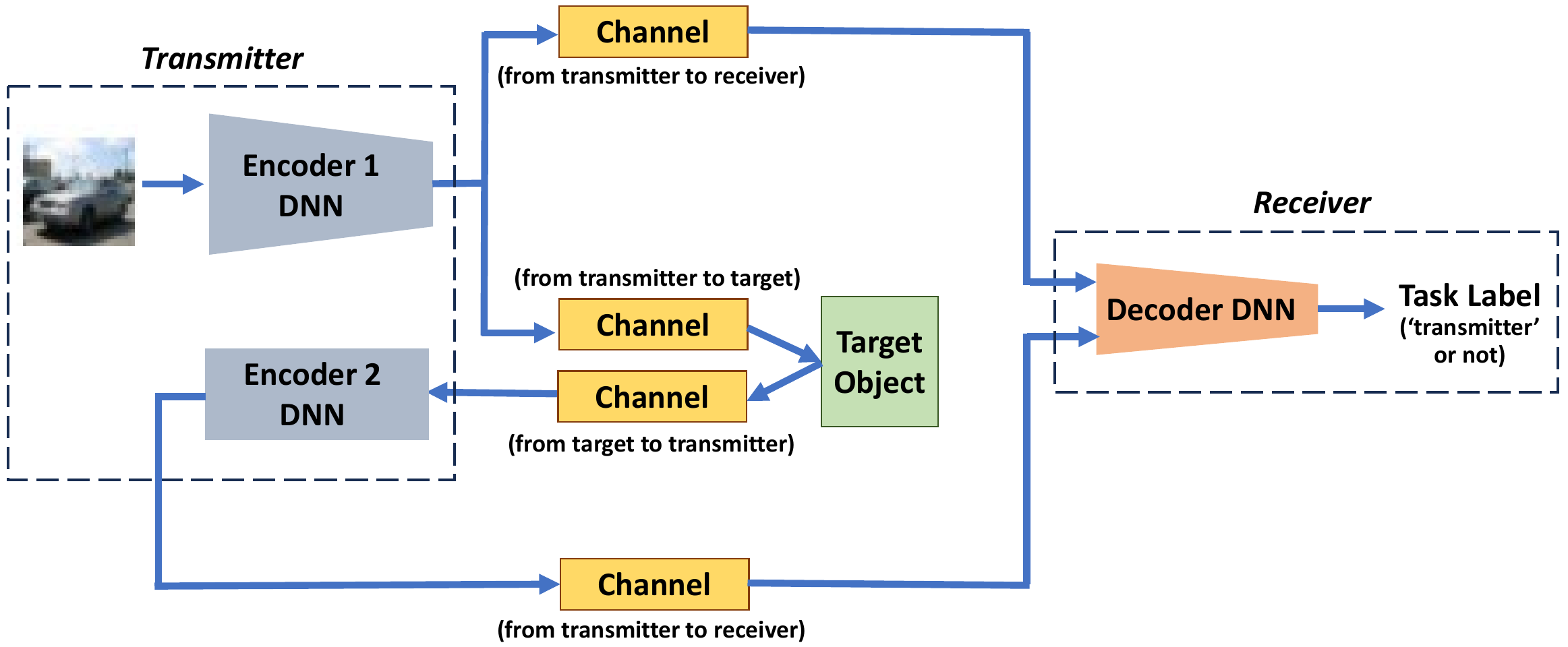}
	\caption{System model for joint sensing and task-oriented communications.}
	\label{fig:JSTOCsystemmodel}
 \vspace{0.2cm}
\end{figure*}

The system model for joint sensing and task-oriented communications, shown in Fig.~\ref{fig:JSTOCsystemmodel}, involves an integrated approach where image data, initially transmitted for object recognition, is utilized for both image classification and target sensing. The following steps are pursued:
\begin{enumerate}
    \item The transmitter (an edge device) captures images.
    \item The transmitter performs task-oriented communications by processing input data samples through an encoder to deliver the reduced amount of necessary information from the image samples to the receiver (fusion center).
    \item Simultaneously, this transmission is used for target sensing by allowing the transmitted signals to interact with the environment and return to the transmitter in a joint sensing and communication framework.
    \item The transmitter performs another round of task-oriented communications by processing the reflected signals through another encoder to deliver the reduced amount of necessary information from the target sensing samples to the receiver (fusion center).
    \item The receiver collects signals from both rounds of task-oriented communication and process the fused data through a decoder to identify potential transmitters. 
\end{enumerate}

\noindent \textbf{Image data at the transmitter.} As the starting point, we consider an edge device that is equipped with a camera to take images. We use the CIFAR-10 dataset as a widely used collection of images designed for object recognition tasks in computer vision. The dataset consists of 60,000 $32\times32$ color images, with each image belonging to one of ten distinct classes. Each image sample has the dimension of $32 \times 32 \times 3 = 3072$. These classes include common objects such as airplanes, automobiles, birds, cats, deer, dogs, frogs, horses, ships, and trucks. Each class contains 6,000 images. The dataset is evenly split into a training set of 50,000 images and a testing set of 10,000 images.

To tailor the CIFAR-10 dataset for our specific task of distinguishing between potential transmitters (vehicles) and environmental objects (animals), we perform a binary classification by grouping the original ten classes into two labels: ``Vehicles" and ``Animals." Specifically, we merge the classes ``airplanes", ``automobiles", ``ships", and ``trucks" into the ``Vehicles" label, encompassing transportation-related man-made objects. Conversely, the classes ``birds," ``cats," ``deer," ``dogs," ``frogs," ``horses," are grouped into the ``Animals" label, representing a diverse set of natural objects within the environment. This binary classification allows us to focus on the distinction between potential transmitters associated with vehicles and other objects present in the surroundings. The deep learning model trained on this modified dataset then aims to efficiently differentiate between these two overarching categories, contributing to the effective identification of transmitters in DSA scenarios.

\noindent \textbf{First round of task-oriented communications.} The edge device that we refer to as the transmitter has the image data samples but they need to be classified at the receiver. We consider task-oriented communications that serves as a resource-efficient alternative to the conventional approach of delivering and reconstructing original data samples at the receiver. The data collected at the transmitter is encoded and selectively transmitted to the receiver. Rather than transmitting the data samples, task-oriented communications involves encoding information in a manner that aligns with the specific goal of the receiver (namely image classification).

An encoder at the transmitter performs joint source coding, channel coding, and modulation to efficiently represent the essential features needed for object classification. This reduced-dimensional information, tailored to the receiver's task of image classification, is transmitted over the channel. This approach significantly minimizes the amount of data that needs to be transmitted while retaining the critical information required for decision-making.

At the receiver, a decoder interprets the received encoded information, extracting the features necessary for image classification. The decoder effectively reconstructs only the information relevant to the task at hand, bypassing the need for reconstructing the entire set of data samples. This targeted communication strategy not only conserves bandwidth but also divides computational load between the transmitter and the receiver, contributing to lower latency and more efficient decision-making. Finally, the transmitter as an edge device cannot be necessarily trusted to complete the task on its own using its encoder alone.

\noindent \textbf{Joint sensing and communications.} This transmission can be also used for the dual purpose of target sensing in the framework of joint (integrated) sensing and communications. The transmitted signals interact with objects in the environment, including potential transmitters. When these signals encounter objects, reflections occur. These reflections carry information about the objects and their surroundings. The reflected signals return to the transmitter. This return path involves the signals bouncing off objects, potentially modifying their characteristics based on the properties of the objects they encounter. In particular, depending on whether the signal is reflected from a vehicle or animal, reflected signals are received at the transmitter with different signal-to-noise ratios (SNRs), in particular smaller sensing SNR for animals compared to vehicles due to more absorption. Note that the sensing SNR incorporates both the effects of channels to and from the target and the effects of reflections from the target depending on the type of target.

\noindent \textbf{Second round of task-oriented communications.} The transmitter sends these reflected target sensing signals to the received by another round of task-oriented communications. For that purpose, a second encoder is trained at the transmitter to process the target sensing data samples before transmitting them to the receiver.

\noindent \textbf{Transmitter identification at the receiver.} The receiver processes multi-modal data from two rounds of task-oriented communications, the first one for transmission of image data and the second one for transmission of target sensing data. For that purpose, the receiver employs a decoder to classify the signals (collected and combined over two rounds of task-oriented communications) into two labels, namely potential transmitter or not. This decoder at the receiver is jointly trained with the two encoders at the transmitter to minimize the loss between the true and predicted labels.   

\noindent \textbf{Channels.} We consider two types of channel models for both communication from the transmitter to receiver as well as for target sensing. First, we consider the AWGN channel model, in which the received signal at the receiver consists of the transmitted signal along with white Gaussian noise. Second, we consider the Rayleigh fading channel, where the transmitted signal undergoes Rayleigh fading, and white Gaussian noise is introduced to the received signal.

Next, we establish how the encoders and the decoders are connected to form the input-output relationships for the underlying DNNs. Suppose that the transmitter ($T$) has $\bm{x}$ as the input (image) data samples. Let $\bm{h}_{ij}^{(k)}$ denote the channel from node $i$ to node $j$ and $\bm{n}_j^{(k)}$ denote the noise at node $j$ at the $k$-th round of task-oriented communications. We assume that the channel and noise distributions are the same over the two rounds of task-oriented communications, while their realizations will change. $T$ encodes $\bm{x}$ as $E_1(\bm{x})$ and transmits it over the channel to the receiver ($R$). The signal that is received by $R$ is given by 
\begin{equation}
\bm{y}_R^{(1)} =  \bm{h}_{TR}^{(1)}  E_1(\bm{x}) + \bm{n}_{R}^{(1)} 
\end{equation}
at the first round of task-oriented communications. In the meantime, the transmitted signal $E_1(\bm{x})$ is reflected off the object ($O$), namely the target, and returns back to $T$ as
\begin{equation}
\bm{y}_T^{(1)} =  \bm{h}_{TOT}^{(1)}  E_1(\bm{x}) + \bm{n}_{T}^{(1)},   
\end{equation}
where $\bm{h}_{TOT}^{(1)}$ is the combined channel from $T$ to $O$ and from $O$ to $T$. Then $T$ encodes $\bm{y}_T^{(1)}$ as $E_2 \bigl(\bm{y}_T^{(1)} \bigl)$ and transmits it to $R$. The signal that is received by $R$ is given by 
\begin{IEEEeqnarray}{rcl}
\bm{y}_R^{(2)} \: & = & \:  \bm{h}_{TR}^{(2)}  E_2 \bigl(\bm{y}_T^{(1)} \bigl) + \bm{n}_{R}^{(2)} \\
 & = & \: \bm{h}_{TR}^{(2)}  E_2 \bigl(  \bm{h}_{TOT}^{(1)}  E_1(\bm{x}) + \bm{n}_{T}^{(1)} \bigl) + \bm{n}_{R}^{(2)}
\end{IEEEeqnarray}
at the second round of task-oriented communications. 

The signals received at the two rounds of task-oriented communications are combined at $R$ as $\bm{y}_R = \bigl\{ \bm{y}_R^{(1)}, \bm{y}_R^{(2)} \bigl\}$ and processed through the decoder $D$ to return the predicted label $\hat{l}$ that is given by
\begin{IEEEeqnarray}{rcl}
\hat{l} \: & = & \:  D\bigl( \bm{y}_R \bigl) \\
& = & \:  D \bigl( \bigl\{ \bm{y}_R^{(1)}, \bm{y}_R^{(2)} \bigl\} \bigl) \\
& = & \: D \bigl( \bigl\{  \bm{h}_{TR}^{(1)}  E_1(\bm{x}) + \bm{n}_{R}^{(1)}   , \nonumber \\  && \:\:\:\: \:\:\:\:\:\: \bm{h}_{TR}^{(2)}  E_2 \bigl(  \bm{h}_{TOT}^{(1)}  E_1(\bm{x}) + \bm{n}_{T}^{(1)} \bigl) + \bm{n}_{R}^{(2)}      \bigl\}     \bigl). \label{eq:E1E2D}
\end{IEEEeqnarray}

As a benchmark for comparison, we consider the case of target sensing only that uses $\bm{y}_R^{(2)}$ as the input to $D$. In that case, $D$ processes $\bm{y}_R = \bigl\{ \bm{y}_R^{(2)} \bigl\}$ to return the predicted label $\hat{l}$ that is given by
\begin{IEEEeqnarray}{rcl}
\hat{l} \: & = & 
\: D \bigl(  \bm{h}_{TR}^{(2)}  E_2 \bigl(  \bm{h}_{TOT}^{(1)}  E_1(\bm{x}) + \bm{n}_{T}^{(1)} \bigl) + \bm{n}_{R}^{(2)}          \bigl). \label{eq:E1E2D_2}
\end{IEEEeqnarray}

The overall DNN structures used in (\ref{eq:E1E2D}) and (\ref{eq:E1E2D_2}) combine the two encoders $E_1$ and $E_2$, and the decoder $D$ along with the channel effects. $E_1$, $E_2$, and $D$ are jointly trained by accounting for all the underlying channel and noise effects $\bm{h}_{TR}^{(1)}, \bm{h}_{TR}^{(2)}, \bm{h}_{TOT}^{(1)}, \bm{n}_{R}^{(1)}, \bm{n}_{T}^{(1)},$ and $\bm{n}_{R}^{(2)}$. In Sec.~\ref{sec:DL}, we will present the DNN architectures for $E_1$, $E_2$ and $D$, and describe how they are trained.

\section{Deep Learning Framework} \label{sec:DL}
\begin{figure*}[h]
\centering 
\includegraphics[width=2\columnwidth]{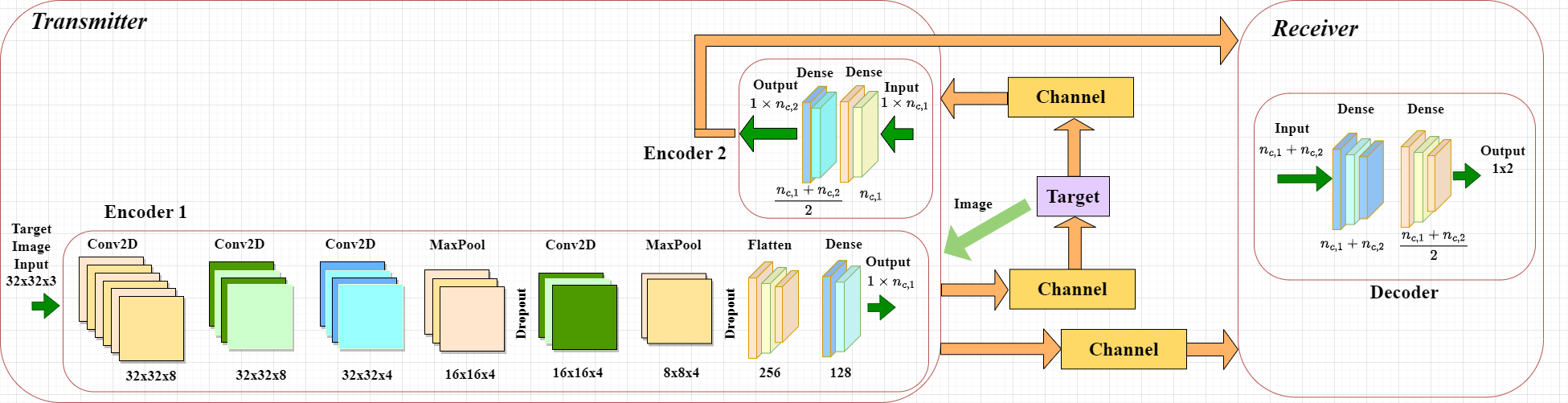}
\caption{Visual description of encoder and decoder DNNs for joint sensing and task-oriented communications.}
\label{fig:DNNs}
\end{figure*}

The DNN architectures used for the two encoders and the decoder are given in Table~\ref{table:DNN}, where $n_{c,1}$ and $n_{c,2}$ are the output sizes of encoders 1 and 2, respectively. The input-output relationships and interactions of DNNs are shown in Fig.~\ref{fig:DNNs} 

We consider a convolutional neural network (CNN) architecture that is separated between two encoders and the decoder. Simpler architectures such as the feedforward neural networks (FNNs) are known to fall short of capturing the complexity of CIFAR-10 dataset. The main building block of the underlying DNNs are the Conv2D, MaxPooling2D, Dropout, Flatten, and Dense layers. 

The Conv2D (2-dimensional convolutional) layer operates by applying a convolution operation to input data in two dimensions. It involves sliding a small filter (also known as a kernel) over the input data, computing the dot product at each position, helping the network capture local patterns and features. This process allows the DNN to learn spatial hierarchies of features in the input. The same set of weights (parameters) in the kernel is shared across the entire input. This sharing reduces the number of trainable parameters, making the model more efficient. Conv2D provides translation invariance, meaning it can recognize patterns regardless of their position in the input. 

The MaxPooling2D layer performs downsampling on two-dimensional input data (feature maps produced by convolutional layers). The main purpose is to reduce the spatial dimensions of the input while retaining the most salient information. MaxPooling2D divides the input feature map into non-overlapping rectangular regions, called pooling windows. Within each pooling window, the maximum value is selected as the representative for that region. These values are then used to create a downsampled version of the input feature map, reducing its spatial dimensions. Similar to convolutional layers, MaxPooling2D provides a degree of translation invariance by focusing on the most prominent features within each pooling window. MaxPooling2D helps reduce the computational complexity of the DNN by discarding less critical spatial information, thus lowering the number of parameters and computations in subsequent layers.

The Flatten layer transforms the high-dimensional data from the preceding layer into a one-dimensional array. This leads to the signals to be transmitted over the channel after going through dense layers to reduce the dimension according to the desired transmitter output size and corresponding number of channel uses. The Dropout layer employs regularization to prevent overfitting. It involves randomly dropping out (setting to zero) a fraction of the neurons during training at each update. The Dense layer in a neural network is a fully connected layer where each neuron receives input from every neuron in the previous layer. It performs a weighted sum of its inputs, adds a bias, and applies an activation function to produce an output. 

We obtain numerical results through Python, and the deep learning model is trained using Keras with the TensorFlow backend. Categorical cross-entropy loss is minimized as the loss between true labels $l$
 and predicted labels $\hat{l}$ (given by (\ref{eq:E1E2D}) and (\ref{eq:E1E2D_2})). The optimizer employed is Adam. The batch size is 64. The number of epochs for training is 5. The resulting DNN structure is trained with 50,000 data samples and tested with 10,000 data samples. 

\begin{table}[h!]
 \captionsetup{justification=centering}
     \caption{Properties of DNN architectures for joint sensing and task-oriented communications.}   
    \label{table:DNN}
	\begin{center}
	\footnotesize
		\begin{tabular}{l|l|l}
			Network & Layer & Properties \\ \hline \hline
			Encoder 1 & Input & size: 32$\times$32$\times$3 \\
& Conv2D & filter size: 8, kernel size: (3,3) \\ & & activation: ReLU \\
& Conv2D & filter size: 4, kernel size: (3,3) \\ & & activation: ReLU \\
& MaxPooling2D & pool size: (2,2) \\
& Dropout & dropout rate: 0.1 \\
& Conv2D & filter size: 4, kernel size: (3,3) \\ & & activation: ReLU \\
& MaxPooling2D & pool size: (2,2) \\
& Dropout & dropout rate: 0.1 \\
& Flatten & -- \\
& Dense & size: 128, activation: ReLU \\
& Dense & size: $n_{c,1}$, activation: Linear \\ \hline
            Encoder 2 & Input & size: $n_{c,1}$ \\ 
		& Dense & size: $n_{c,1}$, activation: ReLU \\
                & Dense & size: $\frac{1}{2} (n_{c,1}+n_{c,2})$, activation: ReLU \\
            & Dense & size: $n_{c,2}$, activation: Linear
            \\ \hline
            Decoder & Input & size: $n_{c,1}+n_{c,2}$ \\ 
			& Dense & size: $n_{c,1}+n_{c,2}$, activation: ReLU \\
                & Dense & size: $\frac{1}{2} (n_{c,1}+n_{c,2})$, activation: ReLU \\
            & Dense & size: $2$, activation: Softmax \\
		\end{tabular}
        
	\end{center}
 \vspace{-0.4cm}
\end{table}

\section{Performance Evaluation} \label{sec:perf}

We consider the following default values of the system parameters. The SNR for communication channel from the transmitter to the receiver is 3dB. The SNR for sensing combines the effect of channels from the transmitter to the target and back from the target to the transmitter as well as the effects of reflections from the targets depending on whether they are vehicles or animals. We set the default value of maximum sensing SNR for vehicles as -3dB and assume that the maximum sensing SNR for animals is 6dB lower. The size of encoder output is 20. Since each image sample has the dimension of $3072$, this corresponds to compression rate of 0.65\%, indicating highly efficient use of communication resources. 

\begin{figure}
\centering
     \begin{subfigure}[b]{0.4\textwidth}
         \centering
         \includegraphics[width=\textwidth]{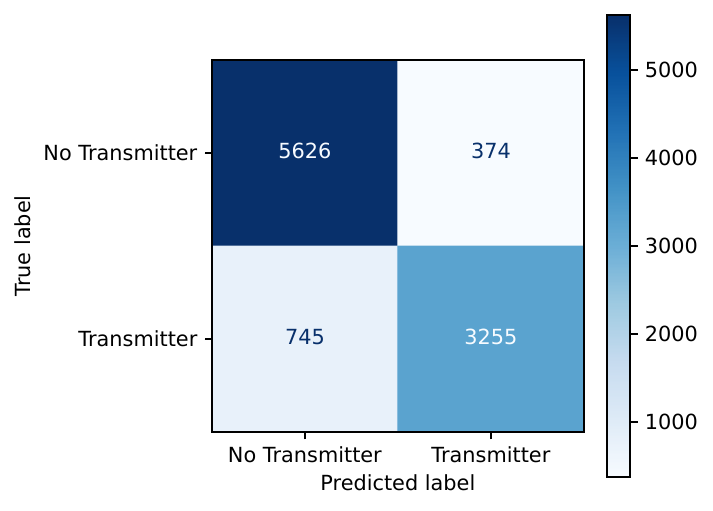}
         \caption{AWGN channel.}
         \label{fig:cmAWGN}
     \end{subfigure}
     \hfill
     \begin{subfigure}[b]{0.4\textwidth}
         \centering
         \includegraphics[width=\textwidth]{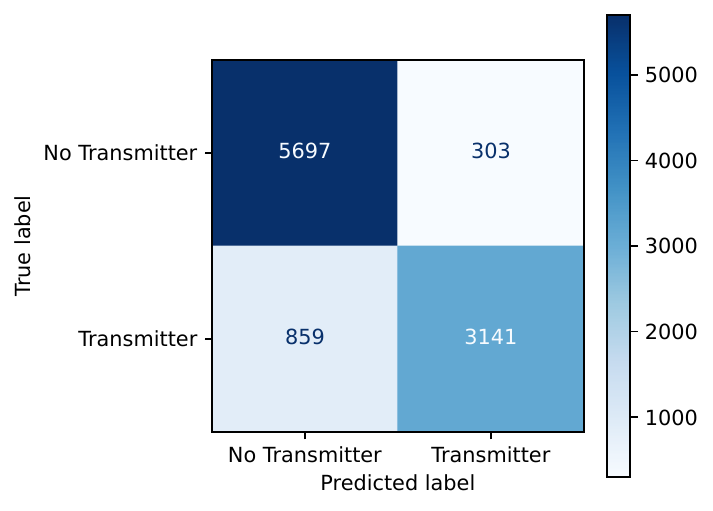}
         \caption{Rayleigh channel.}
         \label{fig:cmRayleigh}
     \end{subfigure}
        \caption{Confusion matrix for default system values.}
        \label{fig:cm}
\end{figure}

Under defaults values of system parameters, the accuracy of joint sensing and task-oriented communications is measured as 0.97 and 0.88 under the AWGN and Rayleigh channels, respectively. Fig.~\ref{fig:cm} shows the confusion matrix under the AWGN and Rayleigh channels for 10,000 test samples. The decoder makes fewer errors in identifying the absence of a transmitter rather than the presence of a transmitter (i.e., false alarm probability is lower than misdetection probability). 

\begin{figure}[h]
	\centering
	\includegraphics[width=\columnwidth]{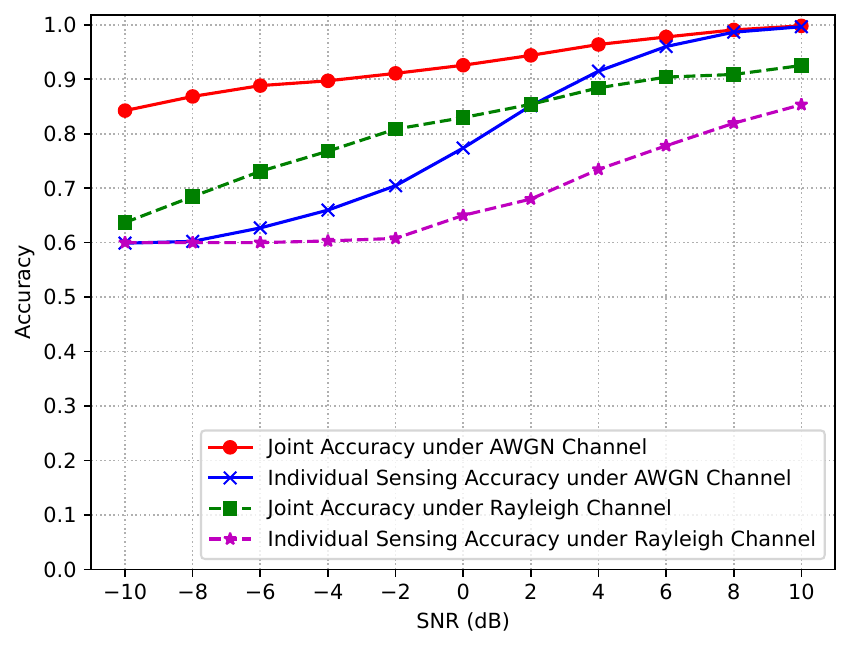}
	\caption{Accuracy vs. communication SNR (dB), when sensing SNR is 6dB lower.}
	\label{fig:NewDyspanJSC_SNR}
 \vspace{0.2cm}
\end{figure}

Next, we vary the system parameters one by one by fixing the rest to default values. Fig.~\ref{fig:NewDyspanJSC_SNR} shows the accuracy as a function of communication SNR (dB), when the sensing SNR is set 6dB lower than the communication SNR. The joint accuracy refers to the accuracy achieved by joint sensing and task-oriented communications, whereas the individual sensing accuracy refers to the accuracy achieved by providing signals reflected off the transmitter as the only input to the decoder. Results show that the joint accuracy is higher than the individual sensing accuracy for AWGN and Rayleigh channels with different communication SNRs. Overall, the accuracy is higher under the AWGN channel compared to the Rayleigh channel. As the communication SNR increases, the accuracy increases in all cases. The individual sensing accuracy catches up with the joint sensing accuracy for high communication SNR such as 10dB under the AWGN channel, whereas the gap is closing slowly with the communication SNR under the Rayleigh channel.

\begin{figure}[h]
	\centering
	\includegraphics[width=\columnwidth]{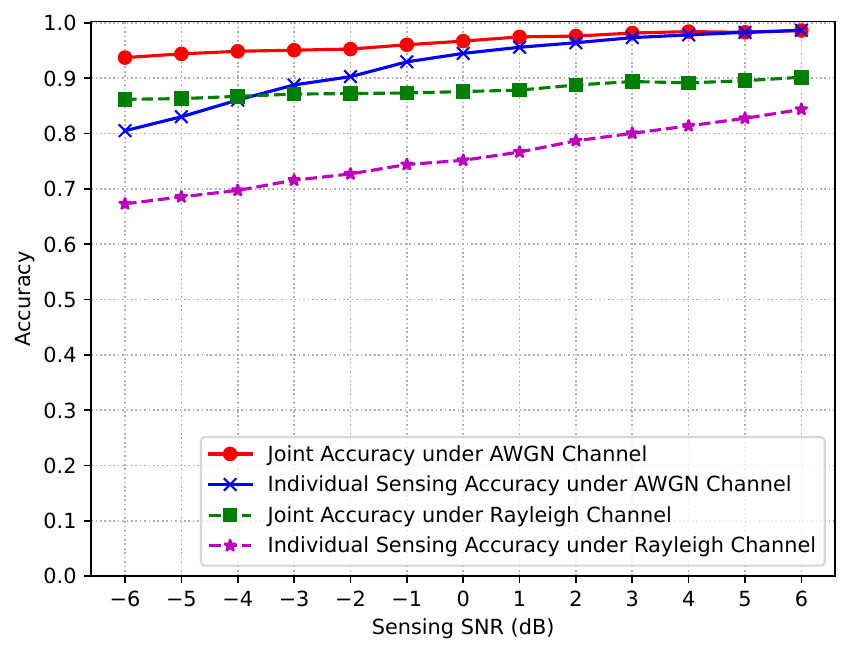}
	\caption{Accuracy vs. sensing SNR (dB), when communication SNR is 3dB.}
	\label{fig:NewDyspanJSC_SNR2}
 \vspace{0.2cm}
\end{figure}

Fig.~\ref{fig:NewDyspanJSC_SNR2} shows the accuracy as a function of sensing SNR (dB), when communication SNR is fixed to 3dB. The effect of sensing SNR is observed more on the individual sensing accuracy compared to the joint accuracy. On the other hand, the gap between the joint sensing accuracy and individual sensing accuracy closes as the sensing SNR increases. In this case, the accuracy is also higher under the AWGN channel compared to the Rayleigh channel. The individual sensing accuracy closes the gap with the joint sensing accuracy as the sensing SNR increases (faster under the AWGN channel than the Rayleigh channel).

\begin{figure}[h]
	\centering
	\includegraphics[width=\columnwidth]{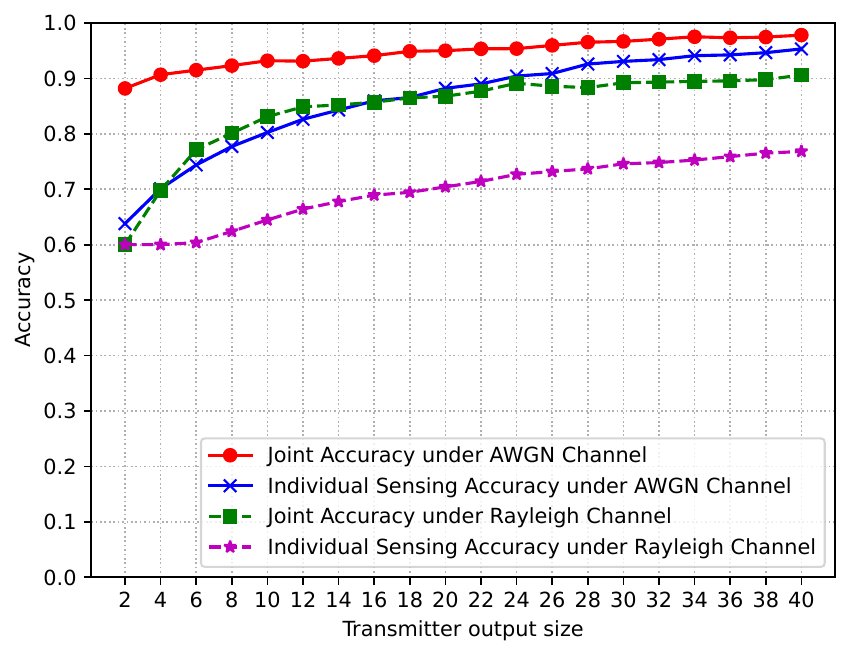}
	\caption{Accuracy vs. transmitter output size.}
	\label{fig:NewDyspanJSC_outputsize}
 \vspace{0.2cm}
\end{figure}

Fig.~\ref{fig:NewDyspanJSC_outputsize} shows the accuracy as a function of transmitter output size for both encoders (namely $n_{c,1} = n_{c,2}$). The joint accuracy remains high even when the transmitter output size is small (which corresponds to high compression of transmitted signals and efficient use of channel resources). In all cases, the accuracy improves with the transmitter output size and starts saturating without the transmitter output size growing significantly. Again, the accuracy is higher under the AWGN channel compared to the Rayleigh channel. While the individual sensing accuracy quickly approaches the joint sensing accuracy as the sensing SNR increases under the AWGN channel, the accuracy gap remains high under the Rayleigh channel.   

\section{Conclusion} \label{sec:Conclusion}
We have presented a novel approach to transmitter identification for DSA by harnessing the power of multi-modal image and target sensing data integration. Focusing on an edge device equipped with a camera to capture images, the system addresses computational constraints and trust issues associated with on-device processing. Through a proposed collaborative framework, the edge device selectively communicates processed information to a trusted receiver acting as a fusion center. Task-oriented communications facilitates efficient transmission of reduced-dimensional information for object classification. Concurrently, the system leverages reflections of transmitted signals to collect target sensing data in a joint sensing and communications framework. The integration of two rounds of encoding at the transmitter and joint training of two encoders at the transmitter and a decoder at the receiver enables seamless fusion of data for image and wireless signal classification. Demonstrating efficacy under AWGN and Rayleigh channel models, the proposed approach attains high accuracy in transmitter identification across diverse channel conditions. The strategic elimination of additional probing signals and the resource-efficient transmission of reduced-dimensional information underscores the practicality of the proposed approach by maintaining low latency. 

\bibliographystyle{IEEEtran}
\bibliography{references}
\end{document}